\documentclass[sigplan,10pt]{acmart} 
\acmSubmissionID{592}

\AtBeginDocument{%
  }

\copyrightyear{2026}
\acmYear{2026}
\setcopyright{cc}
\setcctype{by-nc-nd}
\acmConference[EUROSYS '26]{21st European Conference on Computer Systems}{April 27--30, 2026}{Edinburgh, Scotland Uk}
\acmBooktitle{21st European Conference on Computer Systems (EUROSYS '26), April 27--30, 2026, Edinburgh, Scotland Uk}
\acmDOI{10.1145/3767295.3769358}
\acmISBN{979-8-4007-2212-7/26/04}

\usepackage{xcolor}

\usepackage{algorithm}
\usepackage{algorithmic}
\usepackage{amsmath}

\usepackage{nicefrac}

\usepackage[capitalize]{cleveref} 

\usepackage{xspace}
\usepackage{enumitem}

\usepackage{tcolorbox}

\usepackage{multicol}
\usepackage{multirow}
\usepackage{makecell}
\usepackage{booktabs}

\usepackage{siunitx}

\definecolor{olivegreen}{rgb}{0, 0.6, 0}
\definecolor{redorange}{HTML}{FF5349}
\definecolor{blue(ncs)}{rgb}{0.0, 0.53, 0.74}
\definecolor{navy}{HTML}{273BE2}

\definecolor{black}{HTML}{000000}
\definecolor{white}{HTML}{ffffff}
\definecolor{color1}{HTML}{ACE5EE}
\definecolor{color2}{HTML}{0093AF}
\definecolor{color3}{HTML}{CC0000}
\definecolor{color4}{HTML}{0087BD}
\definecolor{color5}{HTML}{333399}
\definecolor{color6}{HTML}{20B2AA}

\usepackage{colortbl} 

\usepackage{pifont}

\newcommand{\thiswork}{\textsc{FlexiWalker}\xspace}
\newcommand{\thistitle}{FlexiWalker}

\definecolor{redorange}{HTML}{FF5349}

\newcommand{\JL}[1]{{\color{cyan}[\textbf{\sc JLee}: \textit{#1}]}}

\iftrue 
\renewcommand{\JL}[1]{}
\else
\fi

\newcommand{\compilecomp}{\textsc{Flexi-Compiler}\xspace}
\newcommand{\Compilecomp}{\textsc{Flexi-Compiler}\xspace}
\newcommand{\CompileComp}{\textsc{Flexi-Compiler}\xspace}

\newcommand{\runtimecomp}{\textsc{Flexi-Runtime}\xspace}
\newcommand{\Runtimecomp}{\textsc{Flexi-Runtime}\xspace}
\newcommand{\RuntimeComp}{\textsc{Flexi-Runtime}\xspace}

\newcommand{\algocomp}{\textsc{Flexi-Kernel}\xspace}

\newcommand{\rejectoptim}{eRJS\xspace}

\newcommand{\resoptim}{eRVS\xspace}
\newcommand{\Resoptim}{eRVS\xspace}

\newcommand{\thunderrw}{\textsc{ThunderRW}\xspace}
\newcommand{\sowalker}{\textsc{SOWalker}\xspace}
\newcommand{\nextdoor}{\textsc{NextDoor}\xspace}
\newcommand{\csaw}{\textsc{C-SAW}\xspace}
\newcommand{\skywalker}{\textsc{Skywalker}\xspace}
\newcommand{\flowwalker}{\textsc{FlowWalker}\xspace}
\newcommand{\knightking}{\textsc{KnightKing}\xspace}

\newcommand{\naive}{na\"ive\xspace}

\usepackage{tikz}
\newcommand*\circled[1]{\tikz[baseline=(char.base)]{
            \node[shape=circle,draw,inner sep=0.4pt, fill=white, text=black] (char) {#1};}}

\begin{document}

\title[\thiswork]{\thistitle: Extensible GPU Framework for Efficient Dynamic Random Walks with Runtime Adaptation}


\author{Seongyeon Park}
\email{syeonp@snu.ac.kr}
\orcid{0009-0007-3480-6626}
\affiliation{%
  \institution{Seoul National University}
  \city{Seoul}
  \country{South Korea}
}

\author{Jaeyong Song}
\email{jaeyong.song@snu.ac.kr}
\orcid{0000-0001-9976-7487}
\affiliation{%
  \institution{Seoul National University}
  \city{Seoul}
  \country{South Korea}
}

\author{Changmin Shin}
\email{scm8432@snu.ac.kr}
\orcid{0009-0004-8242-8330}
\affiliation{%
  \institution{Seoul National University}
  \city{Seoul}
  \country{South Korea}
}

\author{Sukjin Kim}
\email{iamksj1212@snu.ac.kr}
\orcid{0009-0003-2478-7215}
\affiliation{%
  \institution{Seoul National University}
  \city{Seoul}
  \country{South Korea}
}

\author{Junguk Hong}
\email{junguk16@snu.ac.kr}
\orcid{0009-0001-4004-7714}
\affiliation{%
  \institution{Seoul National University}
  \city{Seoul}
  \country{South Korea}
}

\author{Jinho Lee}
\email{leejinho@snu.ac.kr}
\orcid{0000-0003-4010-6611}
\affiliation{%
  \institution{Seoul National University}
  \city{Seoul}
  \country{South Korea}
}

\renewcommand{\shortauthors}{Park et al.}

\begin{abstract}

Dynamic random walks are fundamental to various graph analysis applications, offering advantages by adapting to evolving graph properties.
Their runtime-dependent transition probabilities break down the pre-computation strategy that underpins most existing CPU and GPU static random walk optimizations.
This leaves practitioners suffering from suboptimal frameworks and having to write hand-tuned kernels that do not adapt to workload diversity.
To handle this issue, we present \emph{\thiswork}, the first GPU framework that delivers efficient, workload-generic support for dynamic random walks.
Our design-space study shows that rejection sampling and reservoir sampling are more suitable than other sampling techniques under massive parallelism.
Thus, we devise (i) new high-performance kernels for them that eliminate global reductions, redundant memory accesses, and random-number generation.
Given the necessity of choosing the best-fitting sampling strategy at runtime, we adopt (ii) a lightweight first-order cost model that selects the faster kernel per node at runtime.
To enhance usability, we introduce (iii) a compile-time component that automatically specializes user-supplied walk logic into optimized building blocks.
On various dynamic random walk workloads with real-world graphs, \thiswork outperforms the best published CPU/GPU baselines by geometric means of 73.44$\times$ and 5.91$\times$, respectively, while successfully executing workloads that prior systems cannot support.
We open-source \thiswork in \url{https://github.com/AIS-SNU/FlexiWalker}.

\end{abstract}




\maketitle

\section{Introduction}

Random walks~\cite{deepwalk, pagerank, ppr, node2vec, metapath, 2ndpagerank, dynnode2vec, reinforced_rw} are pivotal for uncovering structural properties and patterns that drive a broad array of graph‐based applications~\cite{louvainne, infinitewalk, embedding_lstm, evolving, temporal_algo, hetero_gnn, struc2vec, united, asymmetric_proximity, broader_embedding}.
They are particularly effective at distilling core information from large graphs by working on significantly reduced substructures.
Among the many variants, \emph{dynamic random walks}~\cite{node2vec, metapath, 2ndpagerank, dynnode2vec, reinforced_rw} have become especially popular because they capture high‐order patterns and temporal interactions that static walks miss.

Although numerous graph‐processing frameworks exist~\cite{powergraph, graphchi, pregel, ligra} and span diverse platforms~\cite{gunrock, cusha, mosaic, extrav, piccolo, piccolo_cal}, random walks map suboptimally onto them.
In contrast to message‐passing workloads, a random walk must track a \emph{walker} and repeatedly compute a probability distribution over the current node’s neighbors.
This fundamental difference demands a distinct set of optimizations.

A variety of CPU‐ and GPU‐based systems address random walks~\cite{csaw, knightking, thunderrw, drunkardmob, skywalker, nextdoor, noswalker, flashmob, graphwalker, memory-aware-freamework, flowwalker}, but the vast majority target \emph{static} walks.
For static walks, it is profitable to pre‐compute and cache per‐node transition probabilities---an approach that amortizes the dominant cost of constructing such probability distributions~\cite{skywalker, csaw, thunderrw}.

Unfortunately, such reuse techniques do not extend well to \emph{dynamic} walks.
Here, transition probabilities depend on the walker’s history and are revealed only at runtime, forcing each instance to regenerate distributions on demand.
This regeneration incurs repetitive preprocessing or auxiliary data structure construction costs.
Consequently, a framework dedicated to dynamic random walks is needed.

To build such a framework, we first surveyed existing sampling strategies~\cite{alias,fast-its,rejection-sampling,reservoir-sampling} used in prior CPU/GPU systems and assessed their relative compatibility for dynamic walks and massive GPU parallelism.
Among them, we identify rejection sampling and reservoir sampling as the most practical choices for dynamic walks.

\begin{sloppypar}
A closer examination reveals optimization opportunities in the state-of-the-art (SOTA) GPU implementation of both sampling methods.
The leading rejection sampling design~\cite{nextdoor} pays a heavy price for global reductions to compute the maximum transition weight---unnecessary work that we eliminate by computing a bound that does not harm the functionality of sampling.
The SOTA reservoir sampling kernel~\cite{flowwalker} suffers from redundant memory traffic and random‐number generation.
To mitigate these overheads and boost throughput, we adopt an alternative statistical method and a jumping technique for random number generation~\cite{reservoir_expjump, cpu_reservoir_parallel}.
\end{sloppypar}

However, neither of the optimized kernels is a universal winner.
Performance depends on runtime conditions---especially the edge weight distribution, which can shift even within a single walk and can be significantly skewed.
We therefore devise a lightweight first‐order cost model that selects the faster method \emph{per node} at \emph{runtime}.

\begin{sloppypar}
Combining these pieces, we build \thiswork, a GPU framework that delivers efficient dynamic random walks across diverse workloads.
Users write only lightweight workload-specific functions; \compilecomp statically analyzes them to generate specialized building blocks, \runtimecomp chooses the best sampling strategy on the fly, and \algocomp supplies highly optimized kernels.
To the best of our knowledge, \thiswork is the first framework to explore how the optimal sampling strategy varies per sampling instance during dynamic random walk execution.
Such a design is enabled with the tight integration of the compiler component, selection strategy, and kernel optimizations.
\end{sloppypar}

\begin{sloppypar}
We evaluate \thiswork on five widely used dynamic random walk workloads and compare it with representative CPU and GPU baselines.
\thiswork consistently provides optimized solutions where others cannot and achieves geometric mean speedup of 73.44$\times$ and 5.91$\times$ over the best‐performing CPU and GPU baseline cases, respectively.
We release \thiswork in \url{https://github.com/AIS-SNU/FlexiWalker} to facilitate its use.
\end{sloppypar}

\begin{sloppypar}
Our contributions can be summarized as follows:
\begin{itemize}[leftmargin=*]
\item We analyze GPU sampling techniques for \emph{dynamic} random walks and find rejection and reservoir sampling more suitable compared to other techniques when accounting for runtime dynamics and massive parallelism.
\item We optimize both techniques to significantly improve performance by removing global reductions, redundant memory traffic, and random number generation.
\item A lightweight first‐order cost model selects, at runtime and on a per‐node basis, the faster sampling method. The cost model ensures robust performance under highly skewed and time‐varying edge‐weight distributions.
\item By combining compile‐time specialization, adaptive runtime selection, and optimized kernels, \thiswork requires users to supply only workload‐specific logic while automatically delivering state‐of‐the‐art performance.
\item \thiswork surpasses the best‐performing cases of CPU and GPU baselines by geometric means of 73.44$\times$ and 5.91$\times$, respectively, and successfully runs workloads that prior systems could not support efficiently.
\end{itemize}
\end{sloppypar}

\begin{figure}[t]
    \centering
    \includegraphics[width=.95\columnwidth]{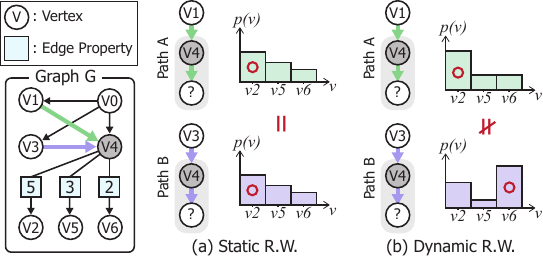}
    \caption{Single step of static and dynamic random walks. 
    }
    \Description{Illustration of a single step in static and dynamic random walks. The static walk follows a fixed transition probability, while the dynamic walk follows varying transition probabilities.}
    \vspace{-1mm}
    \label{fig:rw_gs}
\end{figure}

\section{Background and Analysis}

\subsection{Dynamic Random Walks}
\label{sec:back:walks}

Let $G=(V,E)$ be a directed graph with node set $V$ and edge set $E$.  
For $v \in V$, let $N(v)$ denote the neighbors of $v$, and let $d(v)$ denote its degree.  
For $(v,u) \in E$, $h(v,u) \in \mathbb{R}$ denotes the corresponding edge property weight (i.e., intrinsic edge weight in the graph dataset).

In a random walk, a walker repeatedly selects one of its neighbors (referred to as a \emph{step}) until the predetermined walk length is reached.  
To determine the next node at each step, a probability distribution over the neighbors is required. 
This distribution is constructed from both property weights $h$ and workload-specific weights $w$ assigned according to the target walk algorithm. 
The resulting edge transition weight $\tilde{w}$ and the transition probability of moving from node $v$ to $u$ is defined as: 
\begin{align}
\tilde{w}(v, u) &= w(v, u) \cdot h(v, u), \notag\\
p(u) &= \frac{\tilde{w}(v, u)}{\sum_{t \in N(v)} \tilde{w}(v, t)}
      \quad \text{for } u \in N(v). \label{eq:prob}
\end{align}
Unless specified as property or workload-specific, we use the term \emph{weight} to refer to the final transition weight $\tilde{w}$.

Random walk algorithms are classified as \emph{static random walks} or \emph{dynamic random walks}.  
If the raw edge property values are used directly to build the distribution, i.e., $w(v,u) = 1$, the process is referred to as a static random walk.  
In contrast, if the edge property weights are transformed or adjusted to compute new transition weights based on runtime-specific traits, the process is called a dynamic random walk.

\begin{figure*}[t]
    \centering
    \includegraphics[width=.9\textwidth]{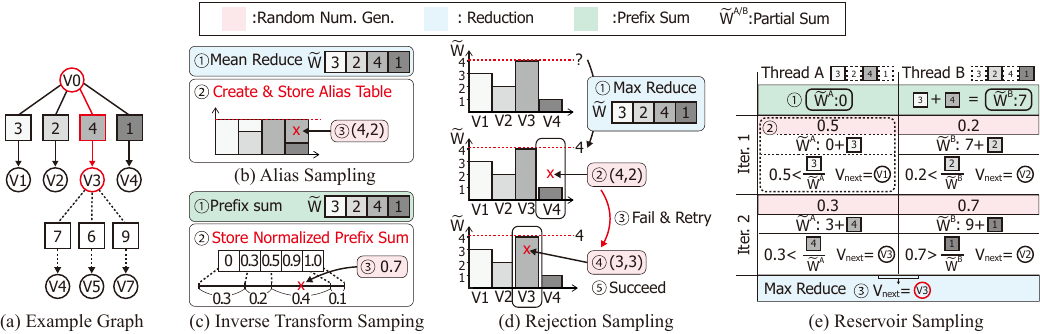}
    \caption{An example of a random walk step and how different sampling methods select the next target node.
    }
    \Description{Illustration of comparing the four sampling methods based on an example graph. Alias and inverse transform sampling require an auxiliary data structure, while rejection and reservoir sampling do not.}
    \vspace{-1mm}
    \label{fig:background}
\end{figure*}

\cref{fig:rw_gs}a shows a step of a static random walk.  
The transition probabilities are identical for walkers that moved from $v_3$ to $v_4$ and from $v_1$ to $v_4$, indicating that the probabilities remain fixed.  
However, in the dynamic random walk illustrated in \cref{fig:rw_gs}b, the probability distribution for the next step differs even though the current node is the same ($v_4$) for both paths~A and~B; the difference arises from their distinct previously visited nodes.

Because the goal of random walks is to extract meaningful node sequences that reflect graph structure and semantics, dynamic random walks are generally more expressive.  
Representative dynamic walks such as Node2Vec, MetaPath, and Second-Order PageRank incorporate graph topology, user inputs, or task-specific functions into the weighting process.

\textit{Node2Vec}~\cite{node2vec} considers the last visited node and updates the transition probability accordingly.  
It modifies $h(v,u)$ by multiplying it with tunable parameters $a$ and $b$.  
Given the current node $v$ and the previously visited node $v'$, the workload-specific weight $w(v, u)$ is determined by $dist(v',u)$, the distance between $v'$ and $u$:
\begin{equation}
\resizebox{0.55\columnwidth}{!}{%
$\begin{aligned}
w(v, u)=
\begin{cases}
\frac{1}{a}, & \text{if } dist(v',u)=0,\\
1,          & \text{if } dist(v',u)=1,\\
\frac{1}{b}, & \text{if } dist(v',u)=2.
\end{cases}
\end{aligned}$\label{eq:node2vec}}
\end{equation}

\textit{MetaPath}~\cite{metapath} operates on graphs with edge labels.  
Given an input schema---an ordered set of labels—the walk must follow this schema while traversing the graph.  
For example, if the schema is $(0,1)$, the first step must follow an edge labeled~0 and the second step an edge labeled~1.  
This effectively sets the values of $w$ to~0 or~1 in accordance with the schema.

\textit{Second-Order PageRank (2\textsuperscript{nd} PR)}~\cite{2ndpagerank}, like Node2Vec, adjusts neighbor weights based on their connection to the previously visited node and also incorporates the degree of said node.  
Given $max_d = \max(d(v), d(v'))$ and a tunable parameter $\gamma$, the workload-specific weight $w(v,u)$ is: 
\begin{equation}
\resizebox{.91\columnwidth}{!}{%
$\begin{aligned}
w(v, u)=
\begin{cases}
\bigl((1-\gamma)/d(v)+\gamma/d(v')\bigr)\cdot max_d, &\text{if } dist(v',u)=1,\\
((1-\gamma)/d(v))\cdot max_d, &\text{otherwise.}
\end{cases}
\end{aligned}$\label{eq:2ndpr}}
\end{equation}

\subsection{Analysis of Sampling Strategies}
\label{sec:back:method}

\begin{sloppypar}
Random walks incur highly irregular memory accesses, and transforming transition weights into a probability distribution demands non-trivial computation.  
The procedure for choosing the next neighbor, including this conversion, is referred to as \emph{sampling}.  
Existing GPU-based implementations~\cite{csaw, flowwalker, skywalker, nextdoor} typically adopt a specific \emph{base sampling method} (e.g., alias sampling~\cite{alias}, inverse transformation sampling~\cite{fast-its}, rejection sampling~\cite{rejection-sampling}, or reservoir sampling~\cite{reservoir-sampling}) and parallelize it to maximize GPU utilization.
\end{sloppypar}

\cref{fig:background} illustrates four commonly used sampling methods on the example graph in \cref{fig:background}a.  
In this example, the current node $v_0$ has four neighbors $v_{1\text{--}4}$ with edge weights $\tilde{w}(v_0,\cdot)=\{3,2,4,1\}$, yielding $p(v_0,\cdot)=\{0.3,0.2,0.4,0.1\}$ for a static random walk.
Suppose the walker randomly selects $v_3$ as the next node ($v_0\!\rightarrow\!v_3$).

\Cref{fig:background}b depicts alias sampling~\cite{alias} (ALS), used in \skywalker~\cite{skywalker}.  
It constructs an auxiliary structure, the alias table, to mitigate random access overhead in weighted sampling.  
ALS distributes weight values across the table as evenly as possible and stores their positions, allowing the next node to be chosen by generating two random numbers that form a 2D lookup coordinate.  
Building the table requires a mean reduction over the edge weights, followed by splitting and redistributing blocks that exceed the mean to other indices.

\begin{sloppypar}
Similarly, inverse transform sampling~\cite{fast-its} (ITS) in \csaw~\cite{csaw} selects the next node with a single random number as shown in \cref{fig:background}c.  
\csaw's ITS constructs a cumulative sum distribution by prefix-sum calculation and normalization. 
Given a random number in $(0,1)$, ITS requires a binary search to find the node that the random number indexes.  
\end{sloppypar}

In contrast, the sampling techniques in \cref{fig:background}d and \cref{fig:background}e do not build any auxiliary data structures.  
\Cref{fig:background}d shows rejection sampling~\cite{rejection-sampling} (RJS), as used in \nextdoor~\cite{nextdoor}.  
Using repetitive trials with $(x,y)$ 2D random coordinates, a node is accepted only if the $y$-value falls within its weight range; otherwise, the trial is rejected.  
Like ALS, RJS needs a reduction to find the maximum weight, but it merely uses it as the upper bound for generating $y$ instead of constructing a table.

\begin{sloppypar}
Finally, \cref{fig:background}e illustrates reservoir sampling~\cite{reservoir-sampling} (RVS), employed by \flowwalker~\cite{flowwalker}, the state-of-the-art GPU framework for dynamic walks.  
RVS visits neighbors sequentially while maintaining a single candidate node: neighbor~$i$ replaces the current candidate if
$u<\tfrac{\tilde{w}_i}{\sum_{k=1}^{i}\tilde{w}_k}$ with $u\!\sim\!\mathrm{Uniform}(0,1)$; otherwise, the candidate is unchanged.
\flowwalker parallelizes RVS by noting that each comparison is independent once the prefix sums $W_i=\sum_{k=1}^{i}\tilde{w}_k$ are available, and uses a max reduction to obtain the final target node.  
Computing the prefix sum and reduction requires inter-thread communication.  
Although generating one random number per neighbor incurs some cost, the absence of intermediate dependencies allows efficient parallelization on GPUs.
For details on the parallelization of RVS, please refer to \cite{flowwalker}.
\end{sloppypar}

\section{Kernel Designs for Efficient Dynamic Random Walks}
\label{sec:kernels}

\subsection{Performance Comparison on Existing Sampling Methods}
\label{sec:motiv:perf}

\begin{figure}
    \centering
    \includegraphics[width=\columnwidth]{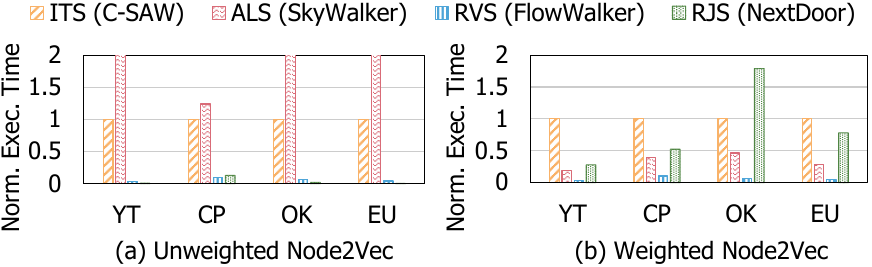}
    \vspace{-4mm}
    \caption{Performance comparison across various sampling methods.  
    Execution time for (a) unweighted Node2Vec and (b) weighted Node2Vec is normalized to ITS (C-SAW).
    }
    \Description{In unweighted Node2Vec, rejection sampling generally outperforms other sampling methods, while in weighted Node2Vec, reservoir sampling outperforms others.}
    \vspace{-2mm}
    \label{fig:motiv:perf}
\end{figure}

Choosing the right sampling method is the critical first step in building an efficient framework for dynamic random walks.  
We believe RJS and RVS are preferable because they do not have the overhead of repetitive auxiliary data structure construction associated with ALS and ITS.
To demonstrate, we measured the runtime of GPU-based random walks that use each sampling method in \cref{fig:motiv:perf} on four datasets~\cite{snap_datasets, law_data1, law_data2}.
We used Node2Vec~\cite{node2vec}, introduced in \cref{sec:back:walks}, as a representative dynamic random walk workload for comparison.

\begin{sloppypar}
\Cref{fig:motiv:perf}a shows the performance for unweighted Node2Vec, where the edge properties are uniform (i.e., $h = 1$), and \cref{fig:motiv:perf}b shows the weighted counterpart.  
As expected, the current baseline implementations of ITS and ALS exhibit significantly longer execution times than the best-performing method, primarily due to the repeated overhead of building auxiliary data structures. 
This confirms that RJS and RVS are preferable for dynamic random walks.  
However, we also observe that the optimal sampling method depends on the environment (i.e., traits of the random walk workload).  
In unweighted Node2Vec (\cref{fig:motiv:perf}a), \nextdoor with RJS overall performs the best, whereas in weighted Node2Vec (\cref{fig:motiv:perf}b), \flowwalker with RVS outperforms the others, and RJS runs much more slowly. 
\end{sloppypar}

These observations led us to select RJS and RVS as the most promising candidates for GPU-based sampling in dynamic random walks. 
In the remainder of this section, we present novel optimizations for each of these methods, which are the first contributions of this paper.

\subsection{Reservoir Sampling Optimization}

\begin{figure}
    \centering
    \vspace{-3mm}
    \includegraphics[width=\columnwidth]{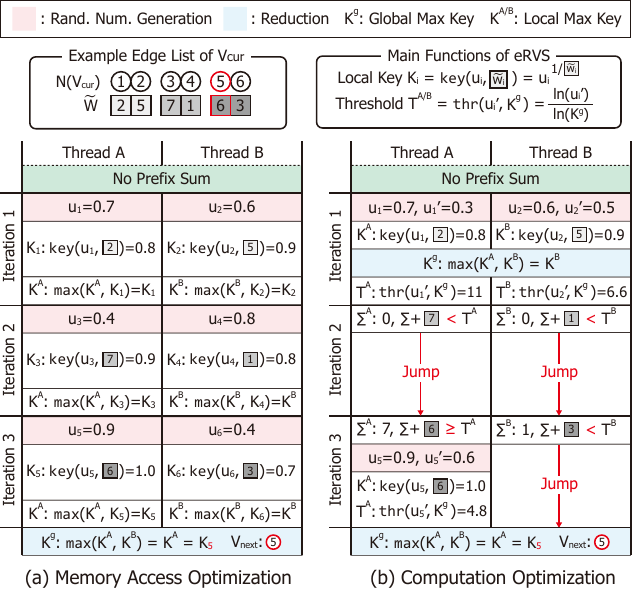}
    \vspace{-6mm}
    \caption{Proposed optimizations of \resoptim.
    }
    \Description{Illustrates how our rejection sampling optimization removes prefix sum and reduces the number of random number generations. For the former, we use an statistically equivalent method, and for the latter, we use a jump technique to eliminate random number generation for certain nodes that meet the jump condition.}
    \vspace{-3mm}
    \label{fig:main:exp_jump}
\end{figure}

Despite the advantages of the RVS approach, the current state-of-the-art GPU-based RVS kernel implementation~\cite{flowwalker} (hereafter called the \emph{baseline RVS kernel}) suffers from two issues: excessive memory access (due to prefix sums) and high computational cost.  
First, the baseline RVS kernel requires the sum of the preceding edge weights for each neighbor visited.
This necessitates full access to the weights and the computation of a prefix sum before sampling. 
Second, the number of random number generations grows exactly in proportion to the current node’s degree, whereas other samplings require far fewer random numbers.  
To tackle these two bottlenecks simultaneously, we propose \resoptim, an enhanced RVS kernel that greatly reduces both memory accesses and random number generations.

\begin{algorithm}[ht]
\caption{RVS with Reduced Memory Access}
\label{alg:exp_rs}
\begin{algorithmic}[1]
 \renewcommand{\algorithmicrequire}{\textbf{Inputs:}}
  \REQUIRE $ $ $v_{cur}$: the current walk vertex (target);\\
  \;\;\;\;\;$N(v_{cur})$: the neighbor sequence of $v_{cur}$;\\
  \;\;\;\;\;$\tilde{w}^{j}_{(\cdot)}$: the dynamic weights of $N(v_{cur})$ in walk step $j$;
 \vspace{2mm}
 \STATE ${k}^{g} \leftarrow -\infty;\quad v_{next} \leftarrow \text{None}$
 \FOR{$i \leftarrow 1$ to $|N(v_{cur})|$}
    \STATE $u_{i} \leftarrow \mathrm{Uniform}(0,1)$ 
    \STATE ${k}_i \leftarrow u_{i}^{\,1/\tilde{w}^{j}_{i}}$
    \STATE \textbf{if} ${k}_i \geq {k}^{g}$ \textbf{then} 
           ${k}^{g} \leftarrow {k}_i$;\quad $v_{next} \leftarrow v_{i}$ 
    \textbf{end if}
 \ENDFOR
 \STATE \textbf{return} $v_{next}$
\end{algorithmic}
\end{algorithm}

To reduce memory accesses, \resoptim first eliminates the prefix sum by adopting a statistically equivalent reservoir-sampling method~\cite{reservoir_expjump, cpu_reservoir_parallel}.  
We illustrate this method in \cref{alg:exp_rs} and its parallel version in \cref{fig:main:exp_jump}a.  
\Cref{alg:exp_rs} converts the sampling step into an \textit{argmax} problem by assigning each neighbor \(i\) a key
\(
   {k}_i = u_i^{\,1/\tilde{w}_i},
\)
where \(u_i \sim \mathrm{Uniform}(0,1)\).  
After the keys are generated, the neighbor with the globally largest key ($k^{g}$) is selected as the next target of the walk.  
Intuitively, each key comprises two factors that determine its magnitude:  
$u_i$ supplies the \textit{randomness} of the walk, and $1/\tilde{w}_i$ acts as a \textit{fixed regulator} of that randomness. 
This strategy removes the need for a prefix sum over the weights, roughly halving the costly memory accesses to the weights.

\Resoptim further reduces the computation by cutting the number of random number generations with the \textit{jump} technique~\cite{devroye1986,reservoir_expjump,cpu_reservoir_parallel}.
Rather than simulating every event with a random key, we exploit the distribution that governs the number of neighbor evaluations required before a candidate update occurs.  
By sampling from this distribution once, we can directly “jump’’ to the neighbor that triggers the update.

According to \cite{reservoir_expjump}, the update to \(v_{\mathrm{next}}\) occurs at neighbor \(s_m\) for which
\begin{align}
  {\textstyle\sum\nolimits_{i=1}^{m-1}} \tilde{w}_{s_i}
  \;<\;
  \nicefrac{\ln(u)}{\ln(k^{g})}
  \;\le\;
  {\textstyle\sum\nolimits_{i=1}^{m}} \tilde{w}_{s_i},
  \label{eq:jump}
\end{align}
where \( \{s_1, s_2, \dots\}\) is the set of assessed neighbors.

\Cref{fig:main:exp_jump}b illustrates the overall parallel jump process.  
To initialize, in the first iteration, each thread first computes a key for one neighbor, as shown in \cref{alg:exp_rs}.  
The resulting keys are reduced to obtain the initial shared global maximum $k^{g}$.  
With this global maximum in hand, each thread generates a threshold $T = \ln(u)/\ln(k^{g})$ and scans its assigned neighbors until the condition in \cref{eq:jump} is satisfied. 
For example, in thread A's second iteration in \cref{fig:main:exp_jump}, the local cumulative weight $\Sigma$ does not satisfy \cref{eq:jump}.
Therefore, it can skip the random number generation and corresponding key generation--in other words, it jumps this iteration.
However, in the third iteration, \cref{eq:jump} is satisfied.
Therefore, the thread computes both the key for the current neighbor and a new threshold for the following iterations.
This overall process is repeated until all neighbors have been scanned and the local max keys are reduced to obtain the next target node.

\subsection{Rejection Sampling Optimization} 

We propose \rejectoptim, a random walk kernel that uses rejection sampling and greatly reduces the memory access bottleneck of the current baseline rejection sampling method.  
Typically, the bottleneck of RJS is similar to that of RVS---full access to the entire transition weight list.  
To minimize the number of retries for each random number generation, RJS sets the range of the generated random numbers to be
\([0,\max(\tilde{w}(v, u))]\) for \(u \in N(v)\).  
For example, in \cref{fig:background}d, the range should be \([0,4]\).
However, this procedure requires a max reduction over all transition weights of the neighbors, which entails full access to the list.
Fortunately, an interesting optimization exists.  
When the maximum transition weight is already known, the max reduction is unnecessary, and sampling involves only iterative random number generation; memory is accessed only for the edge weights selected by the \(x\)-dimension of the random number.

Node2Vec~\cite{node2vec} in an unweighted setting (\(h=1\)) is one such example.  
If \(a=1\) and \(b=0.5\) in \cref{eq:node2vec}, the maximum is automatically 2 and the range is \([0,2]\).  
For other scenarios (e.g., weighted Node2Vec or 2\textsuperscript{nd} PR), however, it is infeasible to precompute the maximum without calculating all edge transition weights.

Instead of obtaining the actual maximum value, \rejectoptim uses a theoretical upper bound computed on the fly.  
The key insight is that exact knowledge of the maximum edge weight is unnecessary; an upper bound that is equal to or larger than the maximum weight suffices.
This allows us to eliminate the max reduction at the cost of potentially more failed sampling attempts caused by additional empty space in the 2D table, while maintaining the functionality of the sampling itself.

We first prove that using an upper bound that is equal to or larger than the maximum weight does not harm the functionality of RJS.
In other words, a neighboring node is sampled with the same probability distribution when using the exact max weight and an upper bound larger than said weight. 
Intuitively, given the $i$th neighbor's sample probability $p(v_i)$ and an upper bound $c$, the relative probability of $v_i$ being sampled over $v_j$ is kept since $\frac{(p(v_i)/c)}{(p(v_j)/c)}=\frac{p(v_i)}{p(v_j)}$. 

Next, we provide a more formal proof for using an upper bound by emulating a random walk sampling instance for the target node $u$. 
We use $\mathcal{N}$ and $\text{deg}$ instead of $N(u)$ and $d(u)$ for readability. 
Given the definition of RJS~\cite{robert1999monte}, we define the following probability mass functions (PMFs) on $\mathcal{N}$:    
\begin{equation}
\resizebox{.8\columnwidth}{!}{
$\begin{aligned}
p&: \mathcal{N} \to (0,1]: \text{target PMF, where } \textstyle\sum_{v\in\mathcal{N}}{p(v)}=1 \\
q&: \mathcal{N} \to (0, 1] : \text{proposal uniform PMF,} \\
& \qquad\qquad\qquad \text{ where } q(v)= 1/\text{deg} \text{ for all } v \in \mathcal{N} \\
c &: \text{any constant such that } c \geq max_{v\in\mathcal{N}}\textstyle\frac{p(v)}{q(v)}.
\end{aligned}$
}
\end{equation}

Using these definitions, the upper bound in \rejectoptim is $\frac{c}{\text{deg}}$, since $p(v){\leq}{c}q(v) = \frac{c}{\text{deg}}$. 
Moreover, the definition of $c$ guarantees that $\frac{p(v)}{{c}q(v)}{\leq}1$ holds for every $v$. 
Note that we chose a uniform PMF for $q$ to simulate the implementation of RJS. 

Using the above, we represent a random walk step as:
\begin{enumerate}[leftmargin=*]
    \item \textit{Propose $X$}: Draw $X\sim q$ (i.e., uniformly from $\mathcal{N}$)
    \item \textit{Draw $U$}: Draw $U\sim Uniform(0,1)$
    \item \textit{Accept/Reject}: \textit{Accept} if $U\leq\frac{p(X)}{{c}q(X)}=\frac{p(X)\cdot\text{deg}}{c}(\leq1)$. 
    Otherwise \textit{reject} and return to 1. The accepted value is the \textit{sample}. 
\end{enumerate}

We can now define the probability of accepting a certain node $v$:
\begin{equation}
\resizebox{0.7\columnwidth}{!}{%
$\begin{aligned}
Pr[\text{sample}=v]&= Pr[X=v~\&~\text{accept}] \\
&=Pr[X=v]{\cdot}Pr[U \leq \textstyle\frac{p(v)}{{c}q(v)}] \\
&=\textstyle\frac{1}{\text{deg}}\cdot\frac{p(v)\cdot\text{deg}}{c}=\frac{p(v)}{c},
\end{aligned}$
}
\end{equation}

and the overall acceptance probability as:
\begin{equation}
\resizebox{.55\columnwidth}{!}{
$\begin{aligned}
Pr[\text{accept}]&=\textstyle\sum_{v\in\mathcal{N}}Pr[\text{sample}=v] \\
&=\textstyle\frac{1}{c}\textstyle\sum_{v\in\mathcal{N}}p(v) = \frac{1}{c}.
\end{aligned}$
}
\end{equation}

Thus, the conditional distribution of acceptance is:
\begin{equation}
\resizebox{.7\columnwidth}{!}{
$\begin{aligned}
    Pr[sample=v|\text{accept}]&=\textstyle\frac{Pr[sample=v]}{Pr[\text{accept}]} \\
    &=\textstyle\frac{p(v)}{c}\div\frac{1}{c}=p(v)
\end{aligned}$
}    
\end{equation}

Hence, the probability distribution of the accepted node is the same as $p$, independent of the constant $c$. 
Therefore, using our insight to optimize RJS is not an approximate solution and does not harm the generality of sampling.

The remaining challenge is \emph{how} to find a sufficient upper bound.  
Ideally, the estimate should satisfy two conditions.  
First, the bound must be close to the actual maximum weight; a bound far from this value enlarges the empty space in the 2D table and increases the expected number of trials before a successful sample.  
To achieve this, we estimate the maximum \emph{per sampling step} rather than using a fixed value for all nodes, because the maximum depends heavily on node- and step-specific factors (e.g., the weight distribution of the current node and the degree of the previously visited node).  
Second, each estimation must be as lightweight as possible, since it will be performed repeatedly.

\begin{figure}
    \centering
    \includegraphics[width= \columnwidth]{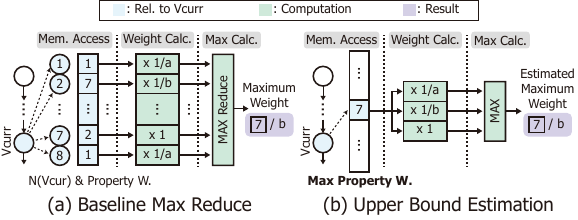}
    \vspace{-7mm}
    \caption{Optimizing rejection sampling with \rejectoptim.
    }
    \Description{The baseline rejection sampling requires full memory access to the edge property weights and corresponding transition weight computation and max reduction. However, our upper bound estimation only requires a max of the property weight and a simple max operation to compute the final estimated maximum weight.}
    \vspace{-2mm}
    \label{fig:main:rjs}
\end{figure}

\rejectoptim proposes to estimate the maximum value by following the variables and features used in the workload's edge weight computation function. 
\cref{fig:main:rjs}a illustrates the baseline process of obtaining the boundary value using weighted Node2Vec. 
While the baseline can obtain the exact boundary value, it requires memory and weight computation per neighbor. 
\cref{fig:main:rjs}b demonstrates how \rejectoptim provides the estimated boundary with only a single memory access and three computations in the case of Node2Vec.
We need to obtain the $max(\tilde{w}(v,u))=max(w(v,u)\cdot h(v,u))$, where we compute its upper bound as $max(w(v,u))\cdot max(h(v,u))$.
The former, $max(w(v,u))$ is obtained from the walk algorithm (e.g., $max(\nicefrac{1}{a}, \nicefrac{1}{b}, 1))$), and the latter, $max(h(v,u))$ can be preprocessed from the input graph dataset.
By finding the max value of both variables, we can obtain the theoretically possible largest boundary per node. 

We note that the aforementioned scheme has two potential pitfalls that hinder the general and efficient usage of \rejectoptim. 
First, if the upper bound is still too far from the actual max weight, its overhead from the increased number of trials can outweigh the benefit of the eliminated max reduction. 
We avoid such cases by using a cost model to select the sampling strategy between rejection and reservoir during runtime, as explained in \cref{sec:main:runtime}.
Second, the estimation process highly depends on the individual random walk algorithm. 
This prohibits the use of a fixed estimation function for all workloads. 
One naive alternative would be to offload the implementation of the estimation function to the users.
However, this would become a tedious process as users have to consider the constants/variables, whether to get the min/max value per variable, and also additional code to generate such values.
We thus propose to automate the process of completing the max estimation pipeline with code analysis and generation in \cref{sec:main:compile}.

\section{\thiswork Framework}

\begin{figure}
    \centering
    \includegraphics[width=\columnwidth]{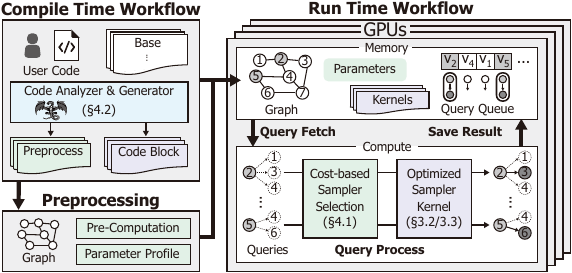}
    \vspace{-5mm}
    \caption{Overview of \thiswork framework. }
    \vspace{-3mm}
    \Description{Illustrates the overview of \thiswork. Given the user input code, our code analyzer generates preprocessing code and other code blocks to plug into \thiswork. The preprocessed parameters are used to select between rejection and reservoir sampling with cost-based sampler selection, and deploy an optimized sampler kernel.}
    \label{fig:overview}
\end{figure}

\thiswork is a flexible framework that automatically runs the user-provided dynamic random walk logic with SOTA performance, using two optimized kernels in \cref{sec:kernels}.
To achieve this, the overall workflow consists of two components: compile-time and runtime workflow.
\cref{fig:overview} provides an overview of those components.

\begin{sloppypar}
At compile time, \compilecomp analyzes the user's workload implementation using LLVM/Clang~\cite{llvm}~(\cref{sec:main:compile}).
Specifically, \compilecomp sweeps through the user code to detect the code structure and features that affect the neighbor sampling probability (e.g., max weight boundary). 
Based on the analysis, \Compilecomp generates human-readable parameters and code blocks, which are plugged into \thiswork's base framework code.
\end{sloppypar}

\begin{sloppypar}
At runtime, \thiswork utilizes the code blocks and parameters from the compile-time workflow.
Optionally, if some parameters (e.g., sampling cost per edge, max/sum of edge property weights) cannot be inferred statically since they are runtime specific, \thiswork deploys a lightweight profiling and preprocessing phase to gather them.
With these, \Runtimecomp concurrently fetches multiple walk queries from the query queue and relevant inputs to sample the next node for each query. 
For every walk step of each query, \Runtimecomp selects the most efficient sampling strategy using a generalizable cost model, informed by profiled data (parameters) and additional runtime-specific characteristics (\cref{sec:main:runtime}). 
The selected strategy is executed via the \rejectoptim and \resoptim kernels (\cref{sec:kernels}), which provide highly optimized kernels tailored to the dynamic nature of the workload.

\end{sloppypar}

\subsection{\RuntimeComp}
\label{sec:main:runtime}

\begin{figure}
    \centering
    \includegraphics[width=\columnwidth]{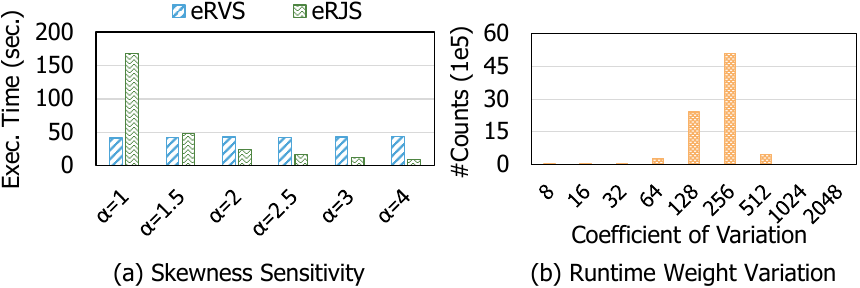}
    \vspace{-5mm}
    \caption{(a) Sensitivity study for reservoir and rejection sampling and (b) runtime weight variation with the EU dataset. 
    }
    \Description{With the EU dataset, we observe that our rejection sampling method's performance varies based on the weight skew, while our rejection sampling shows stable performance. Additionally, using the coefficient of variation for the results of second-order pagerank, we find that the majority of nodes show dynamic sampling behavior during runtime.}
    \label{fig:motiv:powerlaw}
\end{figure}

\textbf{Performance Sensitivity to Weight Distribution.}
In \cref{sec:motiv:perf}, we observed that the best-performing sampling method depends on the environment. 
For a more in-depth analysis of the proposed optimized kernels, we compare the performance of \resoptim and \rejectoptim on weighted Node2Vec over the EU dataset in \cref{fig:motiv:powerlaw}a.
We artificially vary the skewness of the edge property weights $h$ with a power-law distribution.

As shown in the results, \resoptim shows consistent performance regardless of the distribution, as it always performs a single full scan over the entire transition weights. 
On the other hand, the performance of \rejectoptim heavily depends on the skewness, because a single outlier can increase the maximum value, leading to many failed rejection sampling trials.

Additionally, we report the dynamic changes in transition weight distributions during runtime with 2\textsuperscript{nd} PR in \cref{fig:motiv:powerlaw}b. 
We aggregate the sum of edge weights for each node across multiple sampling steps.
For these values, we calculate the coefficient of variation ($CV = std / mean \times 100$) for each node to quantify the relative variability of edge weights during runtime. 
The CVs across all nodes are then used to generate the histogram shown in the figure.

The x-axis indicates the upper bound of each histogram bin (i.e., CV range), and the y-axis denotes the number of nodes that fall into each bin.
The CV metric enables fair comparison across nodes with different mean edge weights; higher CVs indicate greater fluctuations in edge weight sums across steps, highlighting nodes with more dynamic sampling behavior during runtime. 
As shown in the figure, a significant number of nodes exhibit high CV values (toward the right end of the histogram), highlighting substantial runtime variation.
Together, the analysis suggests that selecting an appropriate kernel at runtime is crucial for optimal performance.

\begin{figure}
    \centering
    \includegraphics[width=\columnwidth]{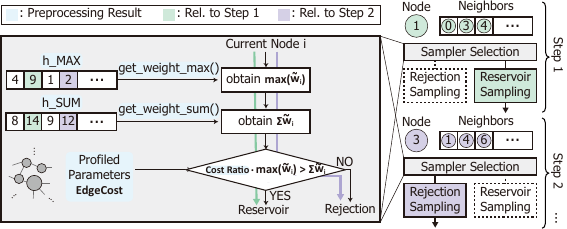}
    \vspace{-5mm}
    \caption{Overall procedure of \runtimecomp. 
    }
    \Description{Our runtime component uses estimated weight max and sum values to compute the cost model. The cost model decides whether to deploy rejection or reservoir sampling for that node.}
    \vspace{-1mm}
    \label{fig:scheme2}
\end{figure}

\textbf{\RuntimeComp Design.}
Based on the above observation, we present \Runtimecomp, a runtime layer to choose the best-fitting sampling strategy for each step.  
To achieve the best performance, we opt to dynamically choose an adequate kernel for each individual step. 
For this purpose, we estimate the overhead of each sampling method by creating a lightweight first-order cost model.

Since both kernels are memory-dominated, our cost model takes into account the number of memory accesses in each method.
From the procedures of \rejectoptim and \resoptim, we can develop a cost model for the number of edge weight accesses for the current target node. 
For \resoptim, the memory cost is simply the degree of the current node $v$,
\begin{equation}
    Cost_{RVS} = EdgeCost_{RVS}\cdot degree,
\end{equation}
where $EdgeCost_{RVS}$ is a sequential memory access cost per edge. 
For \rejectoptim, the model is slightly complicated as we consider the expectation of the sampling trials.
\begin{equation}
    Cost_{RJS} = EdgeCost_{RJS} \cdot degree \cdot {\underset{i}{max}(\tilde{w}_i)}/{\underset{i}{\Sigma}\tilde{w}_i}.
\end{equation}
In this equation, $EdgeCost_{RJS}$ is the random memory access cost per edge. 
The average number of accesses is estimated with the area of the 2D table ($degree \cdot {\underset{i}{max}(\tilde{w}_i)}$) divided by the success space area ($\underset{i}{\Sigma}\tilde{w}_i$).
Comparing the two yields the condition below for preferring \rejectoptim over \resoptim for the current node.
\begin{equation}
    ({EdgeCost_{RJS}}/{EdgeCost_{RVS}}) \cdot \underset{i}{\max}(\tilde{w}_i) < \underset{i}{\Sigma} \tilde{w}_i.
    \label{eq:runtime}
\end{equation}

The system parameter ${EdgeCost_{RJS}}/{EdgeCost_{RVS}}$ is profiled, as detailed in \cref{sec:impl}.
Instead of $\underset{i}{\max}(\tilde{w}_i)$, we use its estimated upper bound due to the \rejectoptim optimization.
Similarly, we use an estimation of ${\sum_i} \tilde{w}_i$ because it dynamically changes, and exact computation is costly.
From \cref{eq:prob}, and applying the linearity of expectation under the assumption that $w$ and $h$ are approximately independent, we estimate:
\begin{equation}
     \underset{i}{\Sigma} \tilde{w}_i = \underset{i}{\Sigma} w_i  h_i \simeq \underset{i}{\Sigma} w_i\cdot E[h]. \label{eq:sum_estimation}
\end{equation}
Both values are obtained through compiler-assisted analysis, as described in \cref{sec:main:compile}.

\subsection{\CompileComp}
\label{sec:main:compile}

We introduce \Compilecomp, a compile-time component for analyzing input functions and generating code blocks to provide an automated pipeline for optimizing dynamic random walk workloads.
To realize this, \Compilecomp analyzes the given workload to determine which optimization techniques are applicable with the relevant parameters, specifically for rejection sampling. 
With the analysis results, \Compilecomp generates code building blocks that integrate into the end-to-end optimized framework. 
In the following, we explain the workflow of \Compilecomp. 

\begin{figure}
    \centering
    \includegraphics[width=\columnwidth]{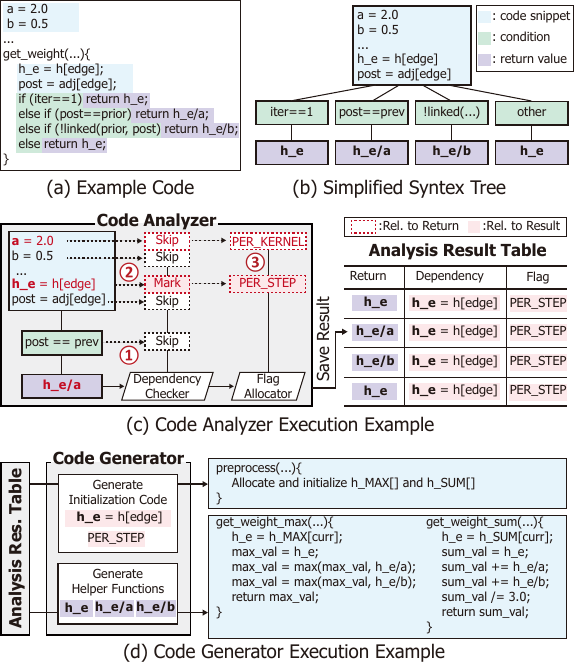}
    \vspace{-5mm}
    \caption{Overall procedure of \compilecomp. 
    }
    \Description{This figure illustrates how our compiler component generates code blocks given the user input code. 
    Using LLVM/Clang, it starts from a leaf of the code syntax tree to store analysis results. 
    These results are then used to generated preprocessing code and helper functions for our optimized rejection sampling.}
    \vspace{-2mm}
    \label{fig:main:compile}
\end{figure}

\begin{sloppypar}
We follow the popular gather-move-update programming model~\cite{thunderrw}, where the core behavior is described in the \texttt{get\_weight()} function as part of the gather phase.
Specifically, for a random walk workload, \compilecomp requires implementing the \texttt{init}, \texttt{get\_weight}, \texttt{update} functions using the CUDA C++ API with the following contents:
\begin{itemize}
    \item \texttt{init}: initializing workload-specific hyperparameters,
    \item \texttt{get\_weight}: computing the current edge's weight,
    \item \texttt{update}: updating query-specific parameters after each walk step.
\end{itemize}
In particular, \texttt{get\_weight()} receives the graph data, workload class, query-specific parameters, and an index of \texttt{get\allowbreak\_weight()}'s target edge, and should return the target edge's weight. 
We present an example user code of weighted Node2Vec in \cref{fig:main:compile}a.
Given fixed hyperparameters (\texttt{a}, \texttt{b}), the current neighbor being explored (\texttt{post}) and its edge property weight (\texttt{h}), weighted Node2Vec checks a number of conditions (green) to compute the final edge weight (purple). 
\end{sloppypar}

With the input code, we deploy \compilecomp's code analyzer, which consists of multiple components for code analysis. 
These include (i) Clang~\cite{llvm} LibTooling-based frontends for source-level abstract syntax tree (AST) analysis, and (ii) LLVM~\cite{llvm} IR-based backends for intermediate representation (IR) analysis.
We integrate both components to enable the analysis of (potentially complex) input code (e.g., control flow checking, data-flow tracking, pointer chasing, alias analysis).
Additionally, we illustrate how \compilecomp views the given input code using a simplified syntax tree in \cref{fig:main:compile}b. 
Starting from the hyperparameters and initial \texttt{get\_weight()} code snippets, each branch represents a possible control flow path, and leaves represent the corresponding return values. 

We demonstrate how \compilecomp's code analyzer operates in \cref{fig:main:compile}c, which comprises a dependency checker and a flag allocator.
The dependency checker tracks expressions that both influence the return value and are necessary for the code generator to function properly. 
Examples include assignment statements or arithmetic operations of the variables that affect the return value. 
The flag allocator detects whether the upper bound estimation for \rejectoptim is applicable to each return variable, and to the entire \texttt{get\_weight()} function. 
Specifically, each flag represents the granularity of weight upper bound estimations needed during runtime, which includes:
\begin{itemize}[leftmargin=*]
    \item \texttt{PER\_KERNEL}: Only a single estimation is required (e.g., unweighted Node2Vec).
    \item \texttt{PER\_STEP}: A boundary estimation per step is required (e.g., weighted Node2Vec).
\end{itemize}
For instance, if the return value uses indexed variables (e.g.,  $h$, the property weights), the flag allocator sets the flag to \texttt{PER\_STEP} as the return value can change per step. 

We demonstrate \thiswork processing an example return value \texttt{h\_e/a} (\cref{eq:node2vec}) as follows:

\begin{enumerate}[leftmargin=*]
    \item[\circled{1}] The dependency checker traverses the syntax tree and encounters a condition/assignment expression for the \texttt{post} variable. These are skipped as they do not influence the return value.  
    \item[\circled{2}] With additional tree traversal, the checker encounters relevant assignment statements for \texttt{h\_e} and \texttt{a}. While we mark \texttt{h\_e} to store it in the result table, \texttt{a} is skipped as it is a workload-specific fixed hyperparameter (i.e., the expression is not necessary for the code generation process).
    \item[\circled{3}] In a similar manner, the flag allocator sets the flag for dependent values. Since \texttt{h\_e} is assigned using an indexed value (\texttt{h}), the expression is flagged with \texttt{PER\_STEP}.
\end{enumerate}

The code analyzer repeats a similar process for each return branch. 
The results are then stored in an analysis result table, which is used in the next code generation step.
It is worth mentioning that the code analyzer analyzes the code in a fine-grained recursive manner in more detail: it recursively identifies all variables per relevant expression and launches different search processes for each variable (e.g., separate analysis processes for \texttt{h\_e} and \texttt{a}).

As illustrated in \cref{fig:main:compile}d, \compilecomp's generator first outputs code snippets for preprocessing the identified relevant indexed values (\texttt{preprocess()}) with the analysis result table. 
The generated preprocessing code allocates and computes the max/sum of the relevant values using templates to create pointers (\texttt{h\_MAX}). 
This is done by attaching pre-defined labels (\texttt{\_MAX}) to the original indexed array (\texttt{h}). 
Next, code blocks are generated to call lightweight GPU reduction kernels for each new pointer.

\begin{sloppypar}

Additionally, \compilecomp's generator outputs helper functions such as \texttt{get\_weight\_max()/\_sum()} in \cref{fig:main:compile}d.
These functions are called before each step to choose the sampling strategy in \runtimecomp. 
\begin{itemize}[leftmargin=*]
    \item The max helper function is generated by adding the dependent code snippets stored in the result table, and using dummy variables (e.g., \texttt{max\_val}) to perform reduction operations for each return value.
    \item The sum helper function estimates the sum of weights with the average of all possible return values following \cref{eq:sum_estimation}. We start by replacing index variables with their new sum pointers (e.g., \texttt{h\_SUM}). We accumulate all possible return values and divide them by the number of unique return values. For cases when \texttt{PER\_KERNEL} is set, we multiply the final average by the degree to emulate the weight sum.  
\end{itemize}
The resulting codes from the generator are integrated into \thiswork, completing the full deployable code base. 
We further discuss the generality of \compilecomp by identifying scenarios where it cannot generate helper functions, and how such cases are addressed in \cref{sec:limit}.
\end{sloppypar}

\section{Implementation}
\label{sec:impl}

\subsection{\RuntimeComp Profiling Kernels.}
To improve the sampling method selection in \runtimecomp, \thiswork deploys a lightweight profiling kernel to estimate the ratio of the overhead per edge weight computation. 
Before sampling, we run two kernels, where each kernel computes the weight value for a fixed ratio of the total nodes and a fixed number of their neighbors using \rejectoptim and \resoptim, respectively.
By running the two lightweight kernels, the profiled ratio can indirectly incorporate hardware-specific features, such as cache hit and capacity.
We design the kernel to only compute weights with a limited number of steps and queries, both of which minimize the profiling time as much as possible.
Moreover, we also observed that limiting the number of computed edge weights allowed estimating the parameters in \cref{eq:runtime} with better performance, compared to simulating the sampling process for each query. 
We report the overhead of the profiling kernel in \cref{sec:eval:overhead}.

\subsection{Concurrent Rejection and Reservoir Sampling.}
\runtimecomp maintains a random walk query for each thread, as rejection sampling requires only a single GPU thread for sampling. 
In contrast, reservoir sampling requires multiple coalesced memory accesses; it is more efficient to use at least 32 threads (i.e., CUDA warp).
To address this gap in the optimal processing unit for each sampling method, we design a runtime kernel that works as follows:

\begin{enumerate}[leftmargin=*]
    \item Each thread within a warp fetches a query job and determines the sampling method with \runtimecomp.
    \item \textit{RJS mode}: After a fixed number of independent rejection sampling trials, threads check whether a thread that requires reservoir sampling exists in the warp with CUDA warp intrinsics (e.g., \texttt{\_\_ballot\_sync()}).
    \begin{sloppypar}
    \begin{enumerate}[leftmargin=*]
        \item \textit{RVS mode}: If at least a single thread uses reservoir sampling, the warp synchronizes to share the target query hyperparameters (e.g., target node, step number) through additional warp intrinsics  (e.g., \texttt{\_\_shfl\allowbreak\_sync()}). The entire warp then concurrently executes reservoir sampling.
        \item If no thread uses reservoir sampling, we continue the process by returning to \textit{RJS mode}.
    \end{enumerate}
    \end{sloppypar}
    \item We continue until all queries in the batch are processed. 
\end{enumerate}

Using this kernel design, we can switch between rejection and reservoir sampling with low overhead and deploy different thread granularities to each sampling method.
Note that due to the implementation choice of switching between \resoptim and \rejectoptim, \compilecomp returns warnings when inter-thread communication is detected in the user's input code.
Examples include warp intrinsics (e.g., \texttt{\_\_ballot\_sync()}) and synchronization functions (e.g., \texttt{\_\_syncwarp()}).

\subsection{Dynamic Query Scheduling.}
\thiswork dynamically retrieves random walk queries from a global queue as individual processing units (i.e., GPU threads) complete their current query.
We found that simply using a global counter representing the number of completed queries and an array of initial nodes was sufficient.
Specifically, we use the global counter as an index to retrieve the initial node of the next target query from the array.
The global counter is incremented atomically to avoid processing units accessing the same query.

\section{Evaluation}
\label{sec:eval}

\subsection{Experimental Setup}
\label{sec:eval:env}

\begin{sloppypar}
To evaluate the performance of \thiswork, we conducted experiments on single- and multi-GPU computing environments with optimized software configurations.
Our hardware setup consists of an AMD EPYC 9124P processor with 16 cores and 32 threads.
The system is equipped with 512GB of DDR5 ECC memory.
For GPU(s), we used up to four NVIDIA A6000 GPUs, each featuring 48GB of VRAM.
On the software side, the experiments were conducted on Ubuntu 22.04.4 LTS.
We utilized CUDA 12.1.1 (driver version 550.54.15) for GPU computations and cuRAND 10.3.2.106 for efficient random number generation.    
\end{sloppypar}

We evaluate \thiswork using diverse real-world graph datasets from various domains, including social networks, citation networks, and web graphs.
The detailed statistics of these datasets are presented in \cref{tab:dataset}.
These datasets vary significantly in size,  
ensuring a comprehensive evaluation across different scales.
Datasets YT, CP, LJ, OK, and FS are obtained from Stanford SNAP \cite{snap_datasets}, while EU, AB, UK, TW, and SK are from the Laboratory for Web Algorithmics (LAW) \cite{law_data1, law_data2}.
This diverse dataset selection allows us to assess the scalability and effectiveness of \thiswork across various graph structures and domains.
Following prior works~\cite{thunderrw, knightking, flowwalker}, we generated random real numbers from $[1,5)$ and random integers from $[0, 4]$ for datasets without edge property weights and edge labels, respectively. 

\begin{table}[t]
    \centering 
    \caption{Real-World Graph Datasets.}
    \vspace{-2mm}
    {
    \resizebox{\columnwidth}{!}
    {
    \setlength{\tabcolsep}{30pt}
    \begin{tabular}{lccc}
        \toprule
     \textbf{Graph} & \textbf{$\#$Vertices} & \textbf{$\#$Edges} \\  
    \midrule
     com-youtube (YT) & 1.1M & 6M \\
    cit-patents (CP)	& 3.8M	& 33M \\
    Livejournal (LJ)	& 4.8M	& 86M \\
    Orkut (OK)	& 3.1M	& 234M \\
    EU-2015 (EU)	& 11M	& 522M \\
    Arabic-2005 (AB)	& 23M	& 1.1B \\
    UK-2005 (UK)	& 39M	& 1.6B \\
    Twitter (TW)	& 42M	& 2.4B \\
    SK-2005 (SK)	& 51M	& 3.6B \\
    Friendster (FS)	& 66M	& 3.6B \\
    \bottomrule
    \end{tabular}
    } 
    }
    \label{tab:dataset}

\end{table}

We chose six baselines~\cite{sowalker, csaw, skywalker, flowwalker, nextdoor, thunderrw}, which are widely known CPU- or GPU-based random walk accelerations.
Two baselines~\cite{thunderrw, sowalker} accelerate random walks using CPUs, while the other four baselines~\cite{nextdoor, skywalker, flowwalker, csaw} employ GPUs for random walk acceleration.

\begin{sloppypar}
\begin{itemize}[leftmargin=*]
    \item \sowalker~\cite{sowalker}: state-of-the-art out-of-core CPU-based dynamic random walk framework utilizing RJS and ITS.
    \item \thunderrw~\cite{thunderrw}: state-of-the-art CPU random walk framework supporting RJS (for unweighted Node2Vec) and ITS for dynamic random walks.
    \item \csaw~\cite{csaw}: dynamic-extended version of ITS-based random walk GPU framework.
    \item \nextdoor~\cite{nextdoor}: RJS-based GPU random walk framework. It partially supports dynamic random walk (for unweighted Node2Vec).
    \begin{sloppypar}
    \item \skywalker~\cite{skywalker}: dynamic-extended ALS-based GPU random walk framework.
    \end{sloppypar}
    \item \flowwalker~\cite{flowwalker}: state-of-the-art GPU-based dynamic random walk framework with RVS.
    \item \thiswork (Proposed): our proposed framework.
\end{itemize}
    
\end{sloppypar}

\begin{sloppypar}
If a baseline does not support a dynamic random walk workload, we faithfully extended it to support it by using the sampling methods employed in the baseline.
For \csaw, as it frequently faces out-of-memory issues and its original open-sourced implementation ignores high-degree nodes with over 90,000 neighbors, we scaled the runtime according to the number of neighbors that it should have considered.
Thus, the actual functionality of \csaw is not exact when facing high-degree nodes, but we included the runtime results of \csaw following the previous work~\cite{skywalker}.
\end{sloppypar}

\begin{sloppypar}
We tested three famous dynamic random walk workloads, Node2Vec~\cite{node2vec}, MetaPath~\cite{metapath}, and Second-Order PageRank (2\textsuperscript{nd} PR)~\cite{2ndpagerank}. 
The method by which each workload obtains dynamic weights is described in \cref{sec:back:walks}.    
In the weighted version of a workload, the property weights $h$ affect the transition weights, while the unweighted version uses only the workload-specific weights $w$ to compute the transition weights. 
Unless specified otherwise, we use \textit{weighted Node2Vec} as the main experiment target.
Additionally, we report out-of-memory and out-of-time (i.e., execution time is longer than 12 hours) with \textit{OOM} and \textit{OOT}, respectively. 
We use the main random walk execution time for evaluation, and report \thiswork's preprocessing/profiling time in \cref{sec:eval:overhead}. 
Following the prior GPU-based SOTA~\cite{flowwalker} on dynamic random walks, we set the number of walk steps to 80 for all workloads except MetaPath. 
For Node2Vec, we set $a = 2.0$ and $b = 0.5$.
For MetaPath, we set the schema to $(0, 1, 2, 3, 4)$ and the depth to $5$.
For 2\textsuperscript{nd} PR, we used $\gamma = 0.2$. 
We created walk queries for every node in the graph. 

\end{sloppypar}

\subsection{Performance Comparison}
\label{sec:eval:perf}

\begin{table*}[t]
    \centering
    \caption{Execution time (ms) comparison with uniform property weight distribution on dynamic random walk workloads.}
    \label{tab:results}
    \vspace{-2mm}
    \includegraphics[width=\textwidth]{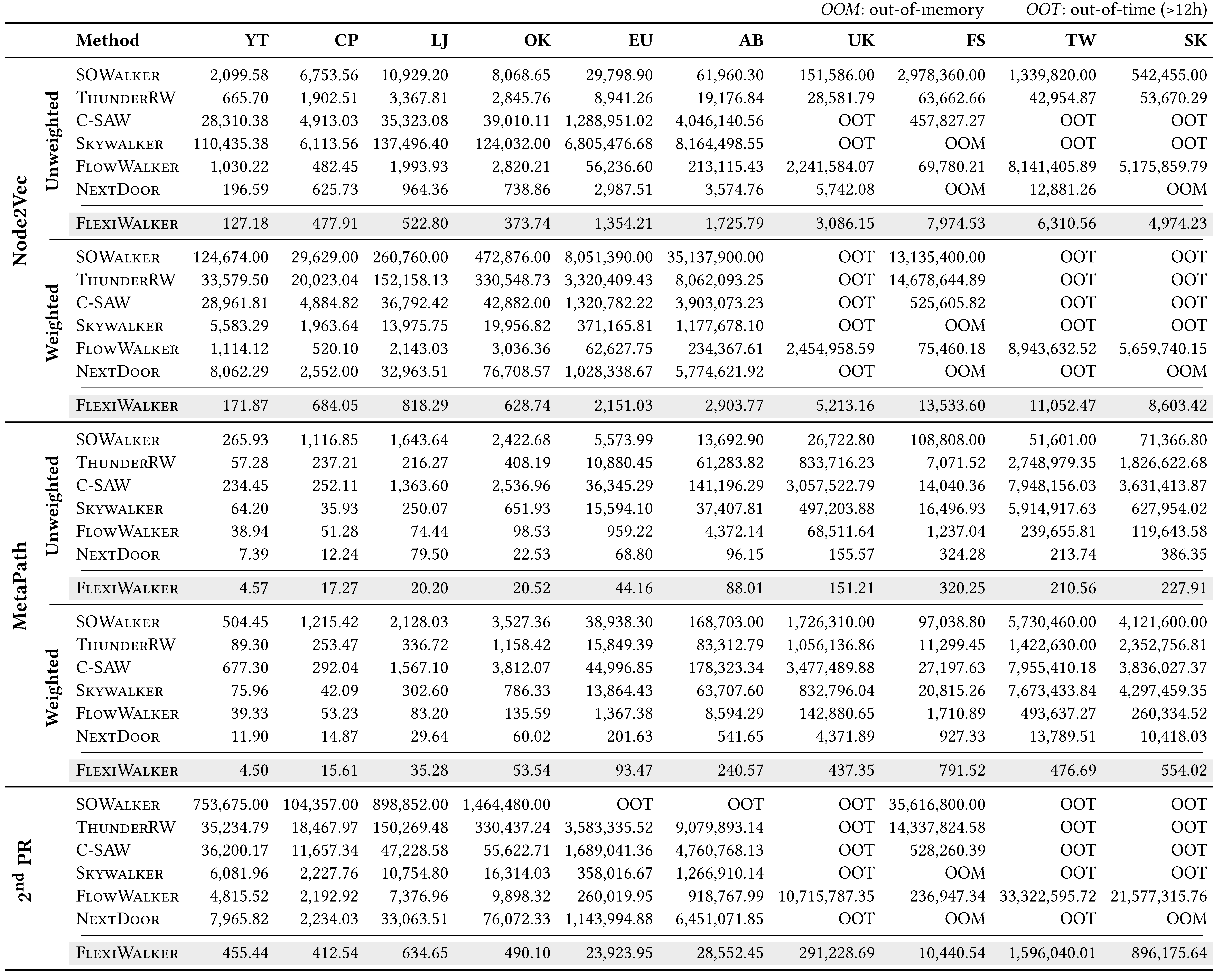}
    \vspace{-4mm}
\end{table*}

\textbf{Uniform Property Weight Distribution.}
\label{sec:eval:perf:uniform}
To benchmark the general dynamic random walk throughput of \thiswork compared with previous frameworks, we tested the uniform edge property weight distribution, which is widely adopted in the evaluation of previous approaches.
\cref{tab:results} shows the execution time of \thiswork and baselines in such weight settings. 
We reported results from a total of five dynamic random walk workloads: (un)weighted Node2Vec, (un)weighted MetaPath, and 2\textsuperscript{nd} PR.

\thiswork generally outperforms other baselines with kernel optimizations and dynamic runtime adaptation for intermediate weight distribution change.
The speedups are significant, providing 4246.71$\times$ maximum and 73.44$\times$ geometric mean speedup over the best-performing cases of CPU baselines and 1040.54$\times$ maximum and 5.91$\times$ geometric mean speedup over the best-performing cases of GPU baselines.
\thiswork achieves more speedup on weighted cases because weighted workloads typically suffer from excessive memory access to obtain max transition weights (i.e., NextDoor) or obtaining weight prefix sum (i.e., FlowWalker). 
For instance, in Node2Vec with AB, \thiswork shows much more speedup in the weighted cases than in the unweighted cases.

Many baselines with inverse transformation sampling and alias sampling often face out-of-time (OOT) on weighted Node2Vec and 2\textsuperscript{nd} PR due to the significant overhead of their sampling methods.
\nextdoor with rejection sampling also suffers from the same issue, since it requires heavy weight max reduction.
On the other hand, \thiswork adopts lightweight and high-throughput \rejectoptim and \resoptim.
As \rejectoptim eliminates weight max reduce and \resoptim minimizes the compute and memory access, \thiswork does not cause OOT in any cases and shows superior throughput in other baselines' OOT cases.
In unweighted Node2Vec with SK, \csaw and \skywalker suffer from OOT, but \thiswork does not, and provides 10.79$\times$ and 1040.54$\times$ speedup over the fastest CPU and GPU baselines, respectively.

\thiswork shows slight slowdowns on some cases with a relatively light workload (i.e., MetaPath).
However, these cases are only confined to small datasets.
For instance, in weighted MetaPath with CP, \thiswork faces 0.05\% slowdown compared to \nextdoor, but this slowdown does not persist on the larger datasets (e.g., \thiswork provides 1.12$\times$ speedup over \nextdoor for OK).
MetaPath has a relatively small number of sampling steps (in this case, five), and hence, the running time is too short on small datasets.
Due to the extremely short running time, the slight overhead of \thiswork causes some slowdowns.

\begin{figure}
    \centering
    \includegraphics[width=\columnwidth]{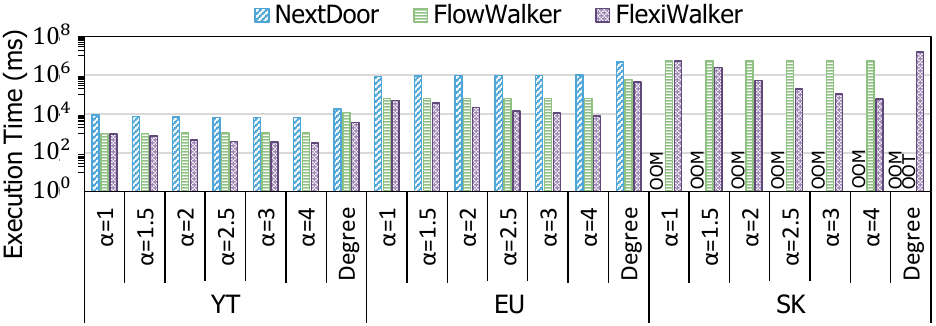}
    \vspace{-5mm}
    \caption{Performance comparison with power-law and degree-based edge property weight distribution.}
    \Description{This figure presents the performance of baselines Nextdoor and Flowwalker against FlexiWalker on power-law and degree-based edge weight distribution. FlexiWalker outperforms the two baselines in all cases.}
    \vspace{-1mm}
    \label{fig:eval:powerlaw}
\end{figure}

\textbf{Power-Law Property Weight Distribution.}
To further test the flexibility of \thiswork on various types of weight distributions, we initialized the edge property weight distribution of graphs to a power-law distribution and benchmarked the dynamic random walk throughput of \thiswork and baselines.
For the initialization, we utilized \texttt{np.random.pareto} function in \texttt{numpy} library.
To emulate the various power-law distributions, we used the Pareto distribution shape value ($\alpha$) from 1.0 to 4.0.
\cref{fig:eval:powerlaw} shows the dynamic random walk execution time of \nextdoor, \flowwalker, and \thiswork for graphs with power-law weight distribution.
We plotted the bars with a log scale, since \nextdoor takes too much time compared to \thiswork.
We included the results of \nextdoor and \flowwalker as they are representative GPU-based rejection sampling and reservoir sampling accelerations and typically showed the highest throughput compared to other baselines in \cref{tab:results}.
Overall, \thiswork provides 26.60$\times$, 4.37$\times$ geometric mean speedup over \nextdoor and \flowwalker, respectively.
\thiswork is more robust on skewed distributions (with lower $\alpha$), thereby showing stable execution time when $\alpha$ is changed.
\nextdoor suffers from the GPU out-of-memory (OOM) issue in the largest SK dataset because it internally uses sorting to exploit memory locality, and thus this sorting requires additional memory usage compared to \thiswork and \flowwalker.

\begin{sloppypar}
\textbf{Degree-Based Property Weight Distribution.}
Lastly, we tested \thiswork and baselines with edge property weights generated with the neighboring node's degree in \cref{fig:eval:powerlaw}.
Due to the nodes with larger degrees having larger edge weights, we observed all baselines taking longer compared to the execution time with uniform and power-law weight distributions. 
Nevertheless, \thiswork outperformed \nextdoor and \flowwalker up to 10.24$\times$ and 3.29$\times$ speedup, respectively. 
Moreover, \thiswork was able to compute all queries in the SK dataset, while both baselines failed either due to OOM or out-of-time (OOT) issues. 
\end{sloppypar}

\subsection{Ablation Studies}

We conducted ablation studies to verify the effect of \thiswork's runtime component, which adaptively selects the sampling kernel and kernel optimizations. 

\begin{figure}
    \centering
    \includegraphics[width=\columnwidth]{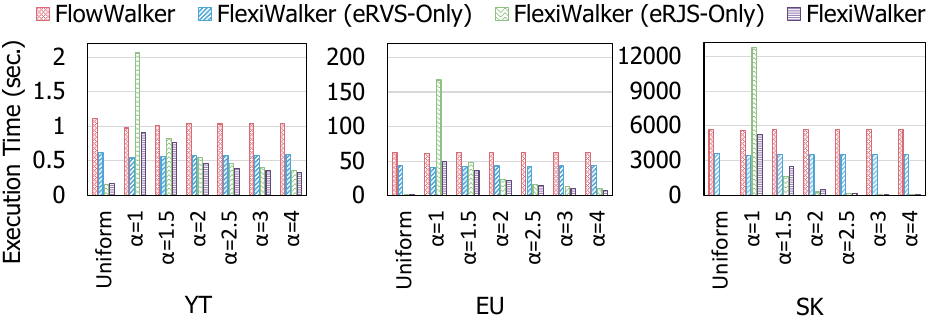}
    \vspace{-5mm}
    \caption{Performance of \thiswork's runtime component with uniform and power-law weight distribution. Other than \flowwalker, all three results are from \thiswork.  
    }
    \Description{This figure demonstrates the performance of FlexiWalker's runtime component. Compared to our rejection and reservoir only methods, our runtime component selects appropriate sampling strategies on-the-fly, providing stable performance benefits.}
    \vspace{-1mm}
    \label{fig:eval:abl:uniform}
\end{figure}

\begin{figure}
    \centering
    \includegraphics[width=\columnwidth]{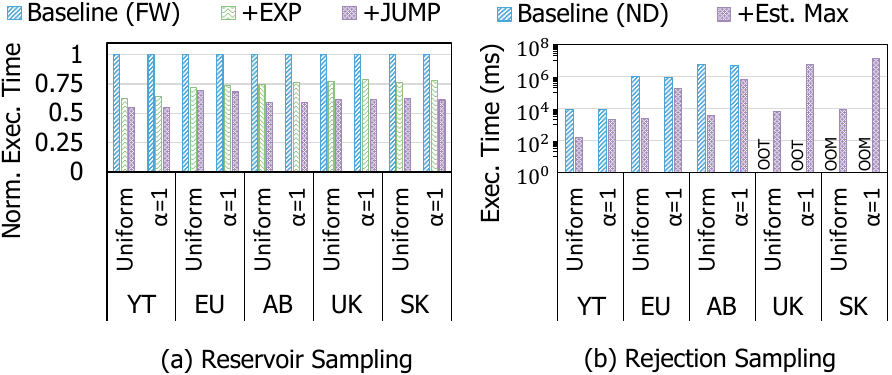}
    \vspace{-2mm}
    \caption{Ablation study of optimization techniques for (a) reservoir sampling and (b) rejection sampling kernels for uniform and skewed ($\alpha=1$) weight distributions. 
    }
    \Description{This figure shows that the optimization techniques used in reservoir sampling and rejection sampling contribute to the final performance of our optimized kernels.}
    \vspace{-2mm}
    \label{fig:eval:abl:algo}
\end{figure}

\textbf{Runtime Component.}
To check the advantage of the runtime component of \thiswork, we compared \thiswork's runtime component with \rejectoptim-only and \resoptim-only versions of \thiswork in \cref{fig:eval:abl:uniform}.
Such single-method versions do not change the sampling kernel during the entire walk procedure. 
We additionally included \flowwalker as a reference, as it is the fastest baseline in weighted Node2Vec.
We tested both graphs with uniform weight distribution (leftmost) and power-law weight distribution (right).
We can observe that \resoptim shows stable performance across varying weight skews, while \rejectoptim's performance degrades as the weight distribution becomes more skewed.
Moreover, for both weight distributions, since the \rejectoptim and \resoptim of \thiswork fail to adapt to node characteristics, \thiswork with the runtime sampling method selection provides up to 3.37$\times$ and 421.56$\times$ speedup over the \rejectoptim and \resoptim versions, respectively.
Note that there are cases where \resoptim/\rejectoptim outperforms the runtime component (e.g., $\alpha=1$ and $\alpha=1.5$ in SK). 
This is due to the runtime component selecting the less optimal sampling strategy for certain instances.
Nevertheless, the runtime component can prevent slowdown when using only either \rejectoptim or \resoptim (e.g., in highly skewed distributions ($\alpha=1$), \rejectoptim is significantly slower than \resoptim).
Overall, this shows the importance of capturing node-specific features during execution in dynamic random walks with the runtime component in \thiswork.

\begin{sloppypar}
\textbf{Kernel Optimizations.}
We further micro-benchmarked the sole effect of kernel optimizations, \rejectoptim and \resoptim, in \cref{fig:eval:abl:algo} with uniform and skewed weight distributions ($\alpha=1$).
In \cref{fig:eval:abl:algo}a, we compared \resoptim with the SOTA reservoir sampling (i.e., \flowwalker).
\Resoptim's advantages compared to \flowwalker are twofold: reduced memory access (EXP) and computation (JUMP).
With only the former, it provides 1.30-1.60$\times$ and 1.27-1.55$\times$ speedup over \flowwalker, and by adding the latter, \thiswork provides 1.44-1.82$\times$ and 1.47-1.81$\times$ speedup over \flowwalker with uniform and skewed weight distribution, respectively.
In \cref{fig:eval:abl:algo}b, we tested \rejectoptim compared to state-of-the-art rejection sampling, \nextdoor.
As \rejectoptim bypasses redundant max reductions with bound estimation, it significantly outperforms \nextdoor from 54.49$\times$ to 1698.35$\times$ in uniform weight distribution.
Similarly, with skewed weight distributions, \rejectoptim outperforms \nextdoor by up to 7.27$\times$.
We believe the relatively lower speedup stems from the skewed distribution, where a large number of candidates are rejected during the sampling process.

\end{sloppypar}

\subsection{Analyses of Sampling Algorithm Selection Strategy}

\begin{figure}
    \centering
    \includegraphics[width=.95\columnwidth]{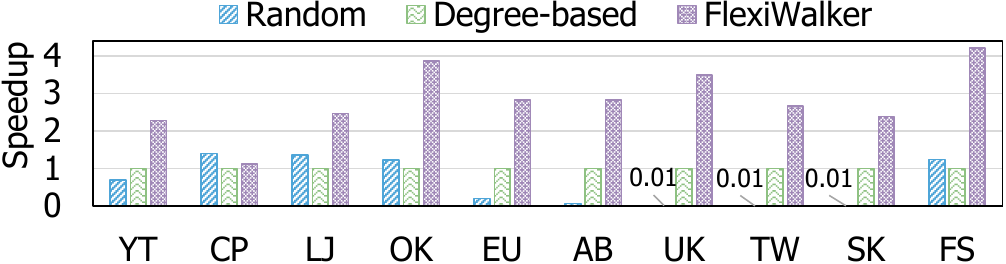}
    \vspace{-1mm}
    \caption{Sensitivity study on sampling method selection strategy (random, degree-based, and \thiswork's cost model).}
    \Description{We compared the performance of different sampling method selection strategies, random and degree-based.
    Compared to the two baseline methods, our cost model shows significant speedup.}
    \vspace{-3mm}
    \label{fig:eval:strat_select}
\end{figure}

\begin{figure}
    \centering
    \includegraphics[width=\columnwidth]{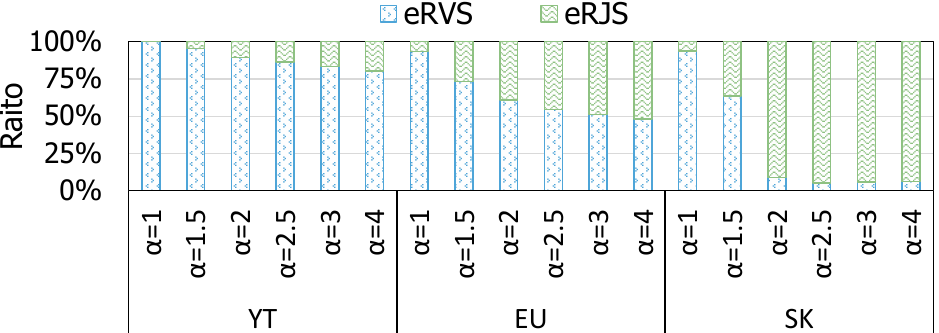}
    \vspace{-3mm}
    \caption{The ratio of chosen sampling method in various power-law weight distributions.}
    \Description{The figure shows how the chosen sampling method between rejection and reservoir sampling changes based on the weight's skewness. For weights with higher skews, reservoir sampling was dominant, while for weights with lower skews, rejection sampling was dominant.}
    \vspace{-3mm}
    \label{fig:eval:select_ratio}
\end{figure}

\textbf{Sensitivity on Selection Strategy.}
To verify whether the sampling algorithm selection engine of \thiswork is adequate, we compared \thiswork with the other two selection strategies, random and degree-based selection in \cref{fig:eval:strat_select}. 
We report the speedup normalized against degree-based selection.
Random selection randomly picks the sampling algorithm between rejection and reservoir sampling.
Degree-based selection employs reservoir sampling when the degree of a node is less than 1K and adopts rejection sampling otherwise, since rejection sampling is more robust for processing high-degree nodes.
\thiswork achieves speedup over random selection and degree-based selection, providing 15.86$\times$ and 2.66$\times$ geometric mean speedup over them.

\textbf{Ratio of Selected Algorithms.}
As \thiswork adaptively selects the sampling algorithm during the walk procedure, it is worth profiling the actual ratio of selection between the two base algorithms, rejection and reservoir samplings.
In \cref{fig:eval:select_ratio}, we profiled the ratio of the chosen sampling method with YT, EU, and SK.
To extract further insight, we tested various power-law Pareto distribution shape values ($\alpha$) ranging from 1.0 to 4.0 (lower values indicate a more skewed distribution).
Since rejection sampling is less robust to skewed distributions than reservoir sampling, rejection sampling is far less selected in skewed distributions with lower $\alpha$.
This shows that \thiswork adequately selects the proper sampling algorithms based on the edge probability distribution.

\subsection{Analysis on Overheads}
\label{sec:eval:overhead}

To verify that \thiswork incurs small additional overhead compared to workload running time, we show the profiling and preprocessing time of \thiswork on all datasets in \cref{tab:preproc}.
Compared to the execution time of weighted Node2Vec, the profiling and preprocessing time are extremely small, requiring only 0.46\%-3.98\% of the total running time.
Also, it is worth noting that those profiled and preprocessed results are reusable per workload and/or graph. 

\begin{table}[t]
  \centering
  \caption{Profile and Preprocessing Time (ms).}
  \label{tab:preproc}
  \vspace{-2mm}
  \resizebox{\columnwidth}{!}{%
    \begin{tabular}{lrrrrrrrrrr}
      \toprule
      \textbf{Time (ms)} & \textbf{YT} & \textbf{CP} & \textbf{LJ} & \textbf{OK} & \textbf{EU} & \textbf{AB} & \textbf{UK} & \textbf{FS} & \textbf{TW} & \textbf{SK} \\
      \midrule
      Profile       & 6.31 & 6.78 & 7.68 & 7.13 & 11.03 & 16.72 & 27.76 & 12.13 & 9.76  & 10.70 \\
      Preproc. & 0.54 & 1.65 & 2.31 & 2.62 & 7.77 & 17.11 & 25.08 & 58.40 & 40.70 & 67.34 \\
      \midrule
      Total         & 6.85 & 8.43 & 9.99 & 9.75 & 18.80 & 33.83 & 52.84 & 70.53 & 50.46 & 78.04 \\
      \bottomrule
    \end{tabular}%
  }
\end{table}

\subsection{Multi-GPU Scalability}

\begin{figure}
    \centering
    \includegraphics[width=.9\columnwidth]{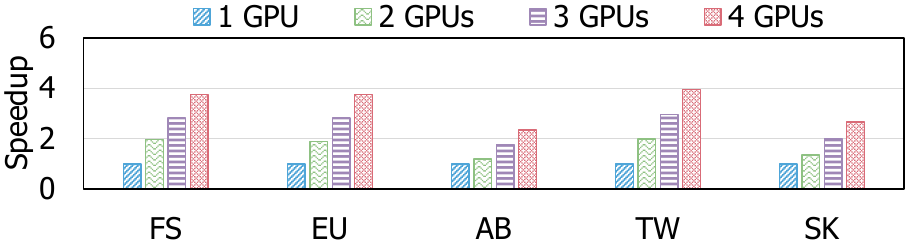}
    \vspace{-2mm}
    \caption{Multi-GPU scalability.}
    \Description{We compared FlexiWalker's performance scalability by increasing the number of GPUs up to four. In general, we observed stable scalability, while a few datasets showed relatively lower speedups due to workload imbalance between GPUs.}
    \label{fig:eval:multigpu}
\end{figure}

We also benchmarked whether \thiswork is scalable with multiple GPUs.
\cref{fig:eval:multigpu} illustrates the scalability of \thiswork as the number of GPUs increases from one to four.
We distributed the queries into GPUs with hash-based index mapping of the starting nodes because \naive range-based index mapping showed lower scalability.
The input graph was duplicated on all GPUs.
Since this query parallelism is highly efficient, \thiswork achieves stable scalability with multiple GPUs, showing 3.23$\times$ geometric mean speedup over the single GPU case using four GPUs.
In AB, \thiswork scales less efficiently due to the remaining workload imbalance among GPUs, but still provides 2.35$\times$ speedup over the single GPU case when utilizing four GPUs.

\subsection{Energy Efficiency}

\begin{figure}
    \centering
    \includegraphics[width=.9\columnwidth]{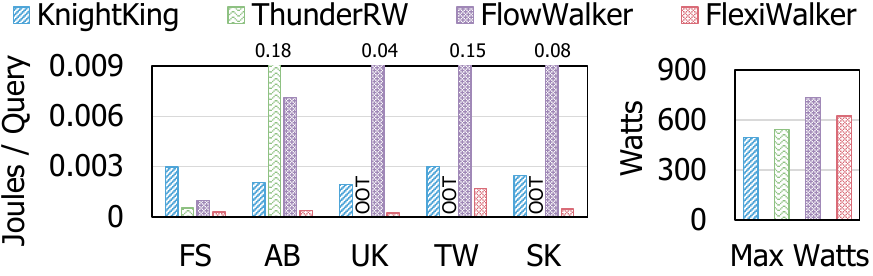}
    \vspace{-1mm}
    \caption{Energy efficiency comparison.}
    \Description{We compare the energy efficiency of CPU baselines (KnightKing, ThunderRW) and GPU baselines (FlowWalker). We observed that KnightKing generally showed the best energy efficiency among baselines, but FlexiWalker was able to outperform KnightKing.}
    \vspace{-1mm}
    \label{fig:eval:energy}
\end{figure}

Lastly, we compared the energy efficiency of \thiswork against CPU and GPU baselines. 
In addition to \thunderrw (CPU) and \flowwalker (GPU), we included another established CPU baseline, \knightking \cite{knightking}. 
We report both the energy efficiency in joules per query and the maximum reported watts of each baseline across all datasets in \cref{fig:eval:energy}. 
We observed \knightking requiring less energy compared to both \thunderrw and \flowwalker, except for FS. 
However, \thiswork was able to achieve the highest energy efficiency with up to 10.15$\times$ less joules per query compared to \knightking. 
Moreover, while \thiswork required higher max watts compared to the CPU baselines, it used 1.18$\times$ less maximum power (watts) than \flowwalker (GPU).

\section{Discussion}

\subsection{Limitations}
\label{sec:limit}

\begin{sloppypar}
We discuss the limitations of \thiswork, focusing on two scenarios where \compilecomp is unable to derive an appropriate upper bound for \rejectoptim.
One scenario is when graph topology-related values are updated during runtime (e.g., updates on edge property weights in dynamic graphs).
Such updates can compromise the accuracy of preprocessed values (e.g., maximum edge property weight), thereby impacting the functionality of \rejectoptim.
Another case stems from complex user input code, where \compilecomp can generate incorrect code for \texttt{get\_weight\_max()/sum()}. 
\Compilecomp thus checks the user code for convoluted loops such as recursive function calls, loops with data-dependent exits, and deeply nested structures. 
When \compilecomp detects such scenarios, it focuses on soundness by safely falling back to \resoptim-only mode.
Nevertheless, \thiswork was able to generate code for all five random walk workloads and utilize \rejectoptim in \cref{sec:eval}. 
We believe \thiswork will be applicable to most dynamic random walk workloads.
\end{sloppypar}

\subsection{Future Extensions}
\label{sec:discuss:extend}

Despite already providing a flexible GPU framework for various dynamic random walk applications, \thiswork has the potential to support additional scenarios. 
First, we can target larger graphs by allocating graph partitions to different GPUs, similar to distributed GNN frameworks~\cite{sancus, granndis}.  
While possible with graph partitioners~\cite{metis} and inter-GPU communication libraries~\cite{nccl, nvshmem}, we expect considerable communication overhead due to the I/O-bound nature of random walks. 
Second, \thiswork can be extended to support dynamic graphs with modules that update the preprocessed values and graph topology. 
\thiswork's design could potentially provide performance benefits for dynamic graphs, especially \runtimecomp, since it can choose different sampling methods per sampling step. 
Finally, \thiswork can support low-precision edge weights to reduce memory bandwidth usage.
To demonstrate, we compared the performance of \thiswork against \flowwalker with weighted Node2Vec using INT8 to store edge property weights (generated with uniform weight distribution). 
\thiswork still outperformed \flowwalker by 27.59$\times$ geometric mean speedup, demonstrating the applicability of \thiswork.

\section{Related Work}

\subsection{CPU-Based Acceleration of Random Walks}

\begin{sloppypar}
Extending static and dynamic random walks to large graphs \cite{graphchi,livejournal,twitter-2010} can be challenging, primarily due to the sheer volume of data and the inherent workload imbalance.
To address such issues, various works target CPUs to optimize random walks~\cite{drunkardmob,bepi,knightking,memory-aware-freamework,soop,graphwalker,noswalker,grasorw,sowalker}. 
DrunkardMob and GraphWalker extend GraphChi for out-of-core simulation~\cite{drunkardmob,graphwalker,graphchi}. 
KnightKing distributes walks with load balancing, using alias sampling for static and rejection sampling for dynamic cases~\cite{knightking}. 
A memory-aware framework~\cite{memory-aware-freamework} analytically selects between different sampling methods but approximates transition weights. 
NosWalker decouples disk I/O from computation for higher throughput~\cite{noswalker}, while GraSorw and SOWalker further optimize out-of-core processing~\cite{grasorw,sowalker}. 
Nonetheless, most schemes still favor static walks or remain bounded by CPU resources.
\end{sloppypar}

\subsection{GPU-Based Acceleration of Random Walks}

Several works~\cite{csaw, nextdoor, skywalker, cowalker, skywalkerplus, flowwalker} have accelerated random walks using GPUs.
C-SAW~\cite{csaw} utilizes the inverse transform sampling algorithm with warp-centric parallelism.
NextDoor~\cite{nextdoor} performs graph sampling based on rejection sampling by gathering workloads with the same target using transit parallelism.
SkyWalker~\cite{skywalker, skywalkerplus} and CoWalker~\cite{cowalker} employ alias sampling on GPUs.
FlowWalker~\cite{flowwalker} is the current state-of-the-art GPU framework for dynamic random walks.
The aforementioned solutions either target static random walks or are burdened by additional performance bottlenecks arising from dynamic random walk traits. 
\thiswork efficiently alleviates such issues using efficient kernels and generalizable optimization techniques.

\begin{sloppypar}
    
\section{Conclusion}
We propose \thiswork, an extensible framework for dynamic random walks.
\thiswork proposes efficient rejection and reservoir sampling kernels, a lightweight cost model for runtime sampling method selection, and compile-time automatic workload optimization and specialization.  
\thiswork achieves 73.44$\times$ and 5.91$\times$ geometric mean speedup over CPU and GPU baselines' best performing cases, respectively.

\end{sloppypar}

\begin{acks}
\begin{sloppypar}
This work was partially supported by 
Institute of Information \& communications Technology Planning \& Evaluation (IITP) 
(RS-2024-00395134, 
RS-2024-00347394, 
RS-2023-00256081, 
RS-2021II211343).   
Part of the infrastructure used in this work was supported by Korea Basic Science Institute (National research Facilities and Equipment Center) grant funded by the Ministry of Science and ICT (No. RS-2025-00564840).
Jinho Lee is the corresponding author.

\end{sloppypar}
\end{acks}

\clearpage

\bibliographystyle{ACM-Reference-Format}
\balance
\bibliography{references}


\end{document}